\documentclass[final]{elsarticle}
\pdfoutput=1
\usepackage[toc,page]{appendix}
\usepackage{lineno}
\usepackage{chemformula}
\usepackage{amsmath,algorithm,algorithmic}
\usepackage{color} 
\newcommand{\dt}{\Delta t}
\newcommand\Da{\rm{Da}}
\newcommand{\erf}{\rm{erf}}
\newcommand{\erfc}{\rm{erfc}}
\title{A second order scheme for a Robin boundary condition in random walk algorithms}

\author[TAU]{Gianluca Boccardo} \ead{boccardo.gianluca@gmail.com}
\author[HUB]{Igor M. Sokolov}
\ead{sokolov@physik.hu-berlin.de}
\author[TAU]{A.~Paster\corref{cor1}}

\address[TAU]{School of Mechanical Engineering, Faculty of Engineering, Tel Aviv University, Israel}
\address[HUB]{Institute of Physics and IRIS Adlershof, Humboldt University Berlin, Germany}
\cortext[cor1]{Corresponding author}

\begin{document}
\begin{abstract}
Random Walk (RW) is a common numerical tool for modeling the Advection-Diffusion equation. In this work, we develop a second order scheme for incorporating a heterogeneous reaction (i.e., a Robin boundary condition) in the RW model. 
 In addition, we apply the approach in two test cases. We compare the second order scheme with the first order one as well as with analytical and other numerical solution. We show that the new scheme can reduce the computational error significantly, relative to the first order scheme. This reduction comes at no additional computational cost.
\end{abstract}

\maketitle
\section{Introduction}

The \textit{advection-diffusion equation} (ADE) is a common tool for describing various transport phenomena.  
In its most basic form (see Eq.~\eqref{eq:ADE}), the equation represents the evolution in time of a scalar quantity of interest due to convection and diffusion. 
The ADE has been employed to describe problems encountered in a variety of scientific fields. For example, transport of contaminants in the environment~\cite{boeker2011book}, chemical reaction engineering~\cite{augier2010}, filtration ~\cite{JCH2017}, semi-conductor physics~\cite{rhoderick1982}, cognitive psychology~\cite{ratcliff2004} and  biological systems~\cite{murray2002book}.
In many applications, the solution of the ADE is obtained by numerical simulations. The numerical approaches can be roughly subdivided into two categories: Eulerian and Lagrangian~\cite{zhang2007}.
In an Eulerian framework, the equation is solved over a fixed grid~\cite{P001,TE2004}~(usually with a finite-volume or finite-elements method). 
In a Lagrangian framework, the unknown variable (e.g. concentration of a chemical species) is modeled by a collection of \textit{particles} which are transported in the domain and may change their properties (e.g. mass) over time ~\cite{majohnson2013,marol2017}. 
A number of hybrid methods, i.e. Eulerian-Lagrangian methods, are also popular~\cite{younes2005,celia1990,neuman1984}. 

Among the different Lagrangian methods, one of the common ones is \textit{particle tracking} (PT). In this approach, the distribution of ``particles" represent (in an approximate manner) the concentration.
The essence of PT is to update the position of the particles every time step via a Langevin equation, i.e. by a combination of a deterministic jump which represents the advection term and a random walk (RW) which represents the diffusion term.

 

Lagrangian approaches in general, and RW schemes in particular, are very popular for solving the ADE, 
due to various advantages over Eulerian schemes.
In Eulerian codes, the concentration is homogenized over a numerical cell, or varies in the cell but in a restricted fashion (e.g. a linear change).
This may smear out sharp concentration fronts, i.e cause a numerical diffusion.
The remedy to this problem is to increase the mesh resolution, but this has a significant computational cost.
Sharp concentration fronts are very common in a wide variety of applications, such as mixing-controlled or very fast chemical reactions, or advection-dominated problems (i.e. problems characterized by high P\'eclet numbers).
In a different vein,  when there is a necessity of employing non-uniform or ``noisy'' initial conditions, RW scheme is quite efficient in describing such conditions~\cite{paster2014connecting}.

Moreover, Lagrangian approaches can tackle problems with an infinite spatial domain.
This is a key difference with respect to Eulerian codes, which need a complete description and discretization of the computational domain.
In the RW codes (being a gridless method) no such restriction applies; it has nonetheless to be noted that the velocity field (i.e., the advection term) has to be known in order to solve the ADE.
In infinite domains, the velocity is known everywhere in the domain only for some special cases where an analytical solution of flow exists.
We shall discuss one such case in the sequel. 

Reactive systems can also be solved by RW codes.
Both homogeneous and heterogeneous reactions can be included in RW codes ~\cite[e.g.,][]{sanchez2015}.
Consider a first order homogeneous reaction (e.g.: a radioactive decay) \ch{A ->[ k ] B} where the reaction rate $r$ depends linearly on the concentration of $A$, i.e. $r=kC_A$. Such case is relatively straightforward to implement in RW~\cite{kinzelbach1988,prickett1981,sherman1986}.
The problem is more complicated when considering a second order homogeneous reaction \ch{A + B ->[ k ] C}, where the reaction rate depends on the concentration of both A and B, i.e. $r=kC_AC_B$ .
This problem was tackled by~\citet{paster2014connecting,paster2013} following the concepts presented in the work of~\citet{benson2008}.
The use of the RW approach for modeling second order reactive systems is especially beneficial when the reaction is diffusion limited and significant concentration gradients develop at the interface between ``islands" of the reactants~\cite{paster2014connecting}.
More recently~\citet{SoleMari2017} presented an approach for including more complicated rate laws, such as Michaelis-Menten, in this RW approach. 

Another significant body of literature was devoted to the inclusion of heterogeneous reactions in RW codes~\cite{agmon1984, schuss2015book, plante2011, prustel2013} and to the implementation of boundary conditions in general.
Some boundary conditions are straightforward to translate from their mathematical definition into RW framework: for example, directly tracking the particles position makes ``inlet'' and ``outlet'' conditions trivial to implement.
Some conditions at solid boundaries are also easy to setup.
An impermeable wall (i.e. where flux is equal to zero, $\partial C/\partial n=0$, $n$ being the normal to the wall) is described by imposing a ``reflection'' rule: particles that cross the boundary are reflected back into the domain.
The more general condition of constant flux (i.e., a Neumann condition, $-D\partial C/\partial n=q(\textbf{x},t)> 0$, where $D$ is the diffusion coefficient) is described in PT by combining the reflection rule and the introduction new particles at the wall~\citep{szymczak2003}.


A more complicated scenario arises when dealing with heterogeneous reactions. These stand as a middle ground between the no-flux condition and the ``perfect-sink'' boundary condition of infinitely fast reaction.
Such heterogeneous reactions can be modeled by a Robin (mixed) boundary condition $-D\frac{\partial C}{\partial n}=kC$ where $n$ is a coordinate normal to the wall.
Even in the Eulerian framework, efforts to accurately employ these conditions are recent~\citep{lin2017high}.
In the current state of the art of RW ~\cite[e.g.,][]{agmon1984,singer2008}, this boundary condition is represented in the algorithm by a certain probability $p$ for the particle that hit the wall to be annihilated and removed from the system.
\citet{agmon1984} derived a first-order accurate expression for this probability, given by $p=p_1=k\sqrt{\frac{\pi D}{\Delta t}}$.
In the case of vanishingly small $\Delta t$, this expression is correct.
In a real application though, there will be a technical constraint on the lower limit for $\Delta t$, dictated by a trade-off between simulation accuracy and computational expense: in these cases, using this expression for reaction probability at the wall could result in an incorrect estimation of particle flux.
In the present work we derive $p_2$, a second-order accurate expression for the reaction probability and discuss the relevance and importance of this result.
We then illustrate the difference between using $p_1$ and $p_2$ by applying them to two test cases and comparing the results with analytical and numerical solutions.

We thus organized this paper as follows.
In the next section, an overview of the theoretical background for the RW methodology in solving the ADE is given, followed by a description of the implementation of the reactive boundary condition in its classical, first-order accurate form.
Then, a theoretical derivation of the higher-order terms is given in Section \ref{sec:convergence_proof}, and specifically the second-order expression which will be used in the remainder of this work.
In Section \ref{sec:exampleApps} two cases will then be shown, where we used a RW code to solve chosen transport problems employing the proposed corrected reaction probability and compared the results with available analytical predictions or grid-converged Eulerian simulations.
In this way, we compare the results of the RW code using its first-order or second-order estimations for the reaction probability, and illustrate the effectiveness of the higher order correction.

\section{Governing equations and the RW methodology} \label{sec:GovEqRWmeth}
\subsection{Governing equations}
For the special case of a constant diffusion coefficient $D$, the advection-diffusion equation is given by

\begin{equation} \label{eq:ADE}
\frac{\partial C}{ \partial t} + {\bf u}\cdot \nabla C =D \nabla^2 C, \quad \textbf{x} \in \Omega
\end{equation}  

\noindent where $C=C(\textbf{x},t)$ is concentration [ML$^{-d}$], $\textbf{u}(\textbf{x},t)$ is the velocity [LT$^{-1}$], and $d=1,2,3$ is the dimension of the domain $\Omega$.
Next, assume that a certain part of the domain boundary $\Gamma_1\in \partial \Omega$, is a non-permeable wall where the perpendicular velocity vanishes, $u_n$=0.
If this wall is reactive, and if the reaction rate is assumed to be linear with the concentration, then the flux of $C$ into the boundary is equal to the rate of reaction.
In this case the boundary condition for $C$ is given by:

\begin{equation}\label{eq:rxnBC}
-D \frac{\partial C}{ \partial n}  = kC, \quad \textbf{x} \in \Gamma_1
\end{equation}  

\noindent where $n$ is the outward coordinate, and $k=k(\textbf{x})\ge 0$ is the reaction coefficient at the boundary. If the boundary is passive, $k=0$; we are mostly concerned here with the reactive case $k>0$.

\subsection{RW scheme}
In the RW numerical scheme, the concentration $C$ is represented by a discrete collection of \textit{particles} in the domain $\Omega$.
Each particle has a specified mass $m_p$.
For the sake of simplicity, and without loss of generality, we shall assume here that all particles have exactly the same mass $m_p=\mathrm{const}$.
Note that these particles are not physical particles.
In contrast, each particle is a point mass (mathematically speaking, a Dirac Delta function).
In other words, the particles can be conceived as a numerical grid which changes over time.  

In RW scheme, a particle's position is updated in each time step by the Langevin equation.
Without loss of generality, we shall consider a one-dimensional problem.
In this problem, the domain is the negative $x$-axis and the governing equation is  

\begin{equation} \label{eq:ADE1d}
\frac{\partial C}{\partial t}+u\frac{\partial C}{\partial x}=D\frac{\partial^2 C}{\partial x^2} , \quad \forall x \le 0,
\end{equation}

\noindent and the boundary condition at $x=0$ is given by

\begin{equation}\label{eq:rxnBC1d}
-D \frac{\partial C}{ \partial x}  = kC \; .
\end{equation}

\noindent The Langevin equation for this one-dimensional  case reads

\begin{equation}\label{eq:Langevin1d}
	x(t+\dt)=x(t)+u\dt+\sqrt{2D\dt}~\xi 
\end{equation}

\noindent where $\dt$ is the time step size, and $\xi$ is a random number with standard normal distribution, i.e. zero mean and unit variance.  

\begin{algorithm} 
	\caption{Random Walk evolution (1D, $x<0$, reactive b.c. at $x=0$)}
\begin{algorithmic} \label{alg:performRW}
	\STATE $t=0$, initialize particles location
	\WHILE {$t<T$}
	\STATE	$t \leftarrow t+\Delta t$, perform RW (Eq.~\eqref{eq:Langevin1d})
	\IF {$x>0$}
	\STATE reflect or annihilate with probability $p$ (see \ref{alg:annihilate_reflected}).
	\ENDIF
	\ENDWHILE
\end{algorithmic}
\end{algorithm}

\subsubsection{Implementation of the reactive BC}
In our numerical scheme, after we moved all the particles by Eq.~\eqref{eq:Langevin1d}, the next step is to implement the boundary condition. Here, a particle with an updated position $x=x(t+\dt)$ such that $x>0$ (i.e., outside $\Omega$),  is reflected back to the domain, to $-x$. 
If the b.c. is prescribed at an arbitrary position $x_0$, the reflected position is $x_0-(x-x_0)=2x_0-x$. 
This process is repeated for each particle, regardless of the whether or not reaction occurs.  
For $k=0$, this assures that the total mass in the domain is preserved. 

For $k>0$, a fraction of the flux at the boundary is consumed by reaction.
In the proposed RW scheme this is implemented by either a complete annihilation of a fraction of the reflected particles, or by a fractional change in the mass of the particles.
Focusing on the first option at present, we need to assign a certain probability $p$ for a reflected particle to be annihilated.
Previous works \cite{agmon1984} have derived this annihilation probability as

\begin{equation}  \label{eq:p_rxn_1}
   p=p_1=k\sqrt{\frac{\pi \dt}{D}}
\end{equation} 
  
\noindent with the apparent requirement that $\dt$ must be small enough such that $p\le1$.

It is relatively straightforward to implement this annihilation probability in the computational code (see Algorithm \ref{alg:annihilate_reflected}).
We note that in RW code with mass-changing particles, the reflected particle will reduce its mass by a relative fraction $p$. 

\begin{algorithm} 
	\caption{Treatment of a reflected particle}
\begin{algorithmic} \label{alg:annihilate_reflected}
	\STATE Generate a random number of uniform distribution $\xi\sim U_{[0,1]}$.
	\IF {$\xi<p$}
	\STATE  The particle is annihilated.
	\ENDIF
\end{algorithmic}
\end{algorithm}

Note that in Eq.~\eqref{eq:p_rxn_1}, at the limit $k\to 0$, all particle are preserved, as expected.
Furthermore, the fraction of annihilated particles is linearly dependent on the reaction rate and the square root of the time step.
The linear dependence on the reaction rate is simple to grasp, since the boundary condition \eqref{eq:rxnBC1d} states that the reaction is linear with $k$.
However the dependence on the square root of $\dt$ is not a straightforward result, and should be explained.
One way to think of this dependence is to look at the case of a single particle that is initially located at some arbitrary position $x=x_0<0$ at $t=0$.
Now, assume this particle performs a random walk for some length of time $T$ and that the time steps taken in this walk have $\dt\ll T$.
Then one can prove that the expected number of hits by a reflective wall located at $x=0$ until $t=T$ is given by

\begin{equation}
   n=\frac{2}{\pi}\sqrt{\frac{T}{\dt}} 
      \left[  \exp(-\chi^2)-\sqrt{\pi}\chi \erfc(\chi)   \right]
\end{equation}

\noindent where $\chi^2=x_0^2/4DT $ is the scaled distance of the particle from the wall. For  $\chi \to 0^+$, the term in the square brackets converges to unity. Hence, for large $T$, $n$ scales like $1/\sqrt{\dt}$ so clearly $p$ must scale like  $\sqrt{\dt}$ to compensate for this.
 
  In the following we prove that \eqref{eq:p_rxn_1} is correct at first order, i.e. it is  $\mathcal{O}(\dt)$, and also provide a second order correction for \eqref{eq:p_rxn_1}. 
 
\section{Convergence of the RW scheme to the reactive BC} \label{sec:convergence_proof}
Our goal here is to prove that the scheme based on Eq.~\eqref{eq:p_rxn_1} converges to the boundary condition \eqref{eq:rxnBC1d} at the limit $\dt\to 0$.
First we note that at a given timestep $\Delta t$, the presence of the boundary affects the particle cloud only at the proximity of the wall.
The characteristic distance of a random walk is  $a'=\sqrt{2D\dt}$, so that the effect of the wall is up to a distance $\mathcal{O}(l)$ from the boundary. Within this infinitesimally short distance from the boundary we assume that the particle distribution follows a stationary smooth concentration given by

\begin{equation} \label{eq:conc_distrib}
	C(x)=C_0+C'x+ \frac{1}{2!} C'' x^2 +\cdots
\end{equation} 

\noindent where $C_0=C(x=0)$, $C'=dC/dx|_{x=0}$, $\dots$ are constants during the specific time step. We further assume the time step of the scheme $\dt$ so small such that $a' \ll C/C'$.
The boundary condition states that the diffusive flux into the boundary equals the rate of reaction at the boundary. In a time step $\dt$ the annihilation rate [ML$^{d-1}$] can be defined as

\begin{equation} \label{eq:annihilation_rate}
	r \dt = -D C' \dt.
\end{equation}

We move now to the particles framework. We first consider a single particle located at $x$ in the beginning of the time step. The probability density function (p.d.f.) of the particle position after $\dt$ is given by the Gaussian

\begin{equation}
	f(x'|x)=\frac{1}{\sqrt{4\pi D \dt}}\exp \left[ - \frac{(x-x')^2}{4D\dt} \right].
\end{equation}

\noindent in an \textit{attempted} step.
If $x'>0$ the particle can either be reflected or annihilated.
The probability of annihilation is thus calculated based on the probability to cross the boundary during the present time step, 

\begin{equation}
	r_1(x)=p \int_{0}^{\infty}f(x'|x)dx'.
\end{equation}

Next we take into account the cloud of particles distributed in $x<0$ with a p.d.f. following \eqref{eq:conc_distrib}. Hence, we take into account all possible values of $x$. The number of particles (or, more appropriately, the mass) in an infinitely small volume $dx$ is $C(x)dx$, and each of them is removed with probability $r_1(x)$, such that the total amount of particles removed in a single time step is 

\begin{equation} \label{eq:No_of_particles_removed}
	r\dt = \int_{-\infty}^{0}C(x) r_1(x) dx = 
	    p  \int_{-\infty}^{0} C(x) \int_{0}^{\infty}f(x'|x)dx' dx ,
\end{equation}

\noindent where we assumed (without loss of generality), that the mass of a single particle is unity.
Performing the integration over $x'$ we get

\begin{equation}
	I=\int_{0}^{\infty}f(x'|x)dx'=\frac{1}{2}+\frac{1}{2} \erf \left(  \frac{x}{\sqrt{4 D\dt}} \right)=
	\frac{1}{2} \erfc \left( - \frac{x}{\sqrt{4 D\dt}} \right)
\end{equation}
such that \eqref{eq:No_of_particles_removed} becomes

\begin{multline} \label{eq:No_of_particles_removed_2}
r\dt = 
\frac{p}{2}   \int_{-\infty}^{0} \left( C_0+C'x+C'' \frac{x^2}{2!}+\dots \right) \erfc \left(  \frac{-x}{\sqrt{4 D\dt}} \right)dx  = \\ 
   a \frac{p}{2\sqrt{\pi}}
        \left( 
                C_0 -    \frac{a\sqrt{\pi}}{4} C' + 
                          \frac{a^2}{6}C'' - \cdots
        \right)
\end{multline}

\noindent where $a^2=4D\dt =(2a')^2$.

\noindent Using \eqref{eq:annihilation_rate} this yields 

\begin{equation}\label{eq:p2}
  p_2=\frac{p_1}{1+p_1/2+\epsilon}
\end{equation}

\noindent where $p_1$ is the first order annihilation probability defined in \eqref{eq:p_rxn_1}, i.e. $p_1=k\sqrt{\frac{\pi \dt}{D}}$ and 

\begin{equation}
\epsilon=\frac{a^2} {6} \frac{C''}{C_0} + \mathcal{O}(a^3)
\end{equation}

\noindent is the sum of the 3rd order and higher order terms.
Note that $\epsilon=\mathcal{O}(\Delta t)$, i.e. it is of order $\sqrt{\dt}$ relative to $p_1$.
The term $\epsilon$ can also be approximated only when $C''$ is known.
In 1D problems, the calculation of $C''$ from particle locations can be done by fitting a 3rd order polynomial to the experimental CDF of the particles close to $x=0$, e.g.: $-3a'<x<0$.
While this can be done in principle, it is cumbersome and may introduce numerical noise into the computation.  

The result~\eqref{eq:p2} is second order accurate.
Dropping $\epsilon \ll \frac{p_1}{2}$ in~\eqref{eq:p2} we get 

\begin{equation}\label{eq:p2_no_eps}
p_2=\frac{p_1}{1+p_1/2},
\end{equation}

i.e., the probability of annihilation $p_2$ is smaller than $p_1$.
For example, when $p_1=0.2$, $p_2 \approx 0.182$ and the reaction rate decreases by $\sim$ 10\%, a rather significant change.
Clearly, when $p_1 \ll 1$, the ratio $p_1/p_2 \rightarrow 1$ and replacing $p_1$ by $p_2$ has a negligible effect.
However, the computational cost involved in having $\Delta t$ small enough such that $p_1 \ll 1$ may be prohibitive.
Then, replacing $p_1$ by $p_2$ has the potential to increase the accuracy of the simulation without any additional computational cost.
To illustrate this, in the following section we will compare the use of $p_1$ \eqref{eq:p_rxn_1} and $p_2$ \eqref{eq:p2_no_eps} in RW codes with analytical and numerical solutions. 

\section{Example applications} \label{sec:exampleApps}
In this section we will explore two different  applications where solution is obtained by RW.  We focus on the treatment of the reactive boundary condition and the accuracy of our proposed methodology.
First, we will consider a simple 1D transient, pure diffusion case where we will compare our simulation results to a known analytical solution.
Then, results of 2D advection-diffusion simulations will be shown, together with a comparison with the results of Eulerian simulations for equivalent setups.

\subsection*{One-dimensional transient pure diffusion}
\begin{figure} 
	\begin{center}
		\includegraphics[width=1\textwidth]{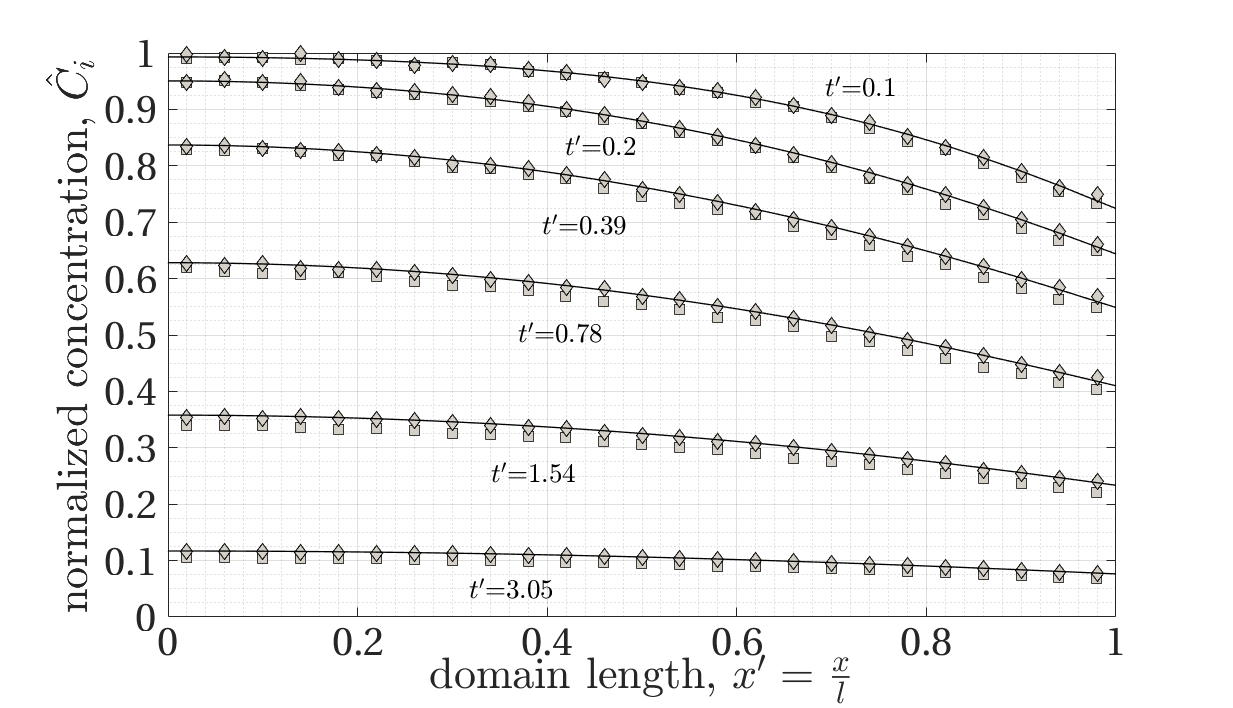}
		\caption{Time evolution of the concentration in the 1D pure diffusion problem, for Da=1. Due to symmetry, only half of the domain is shown.
			The analytical solution (continuous line) is compared with the random walk simulations employing the first and the second order approximations for $p$ (squares and diamonds, respectively). Here, $\Delta t'=5\times10^{-3}$, $N_{p,0}=5\times10^6$, $p_1=0.1253$, $p_2=0.1179$.}
		\label{fig:1D-CTimeSpace}
	\end{center}
\end{figure}

We start by considering the rather simple case of a one-dimensional finite domain of length $2l$.
The governing equation for this problem is

\begin{equation}\label{eq:PureDiff1d}
\frac{\partial C}{\partial t}=D\frac{\partial^2 C}{\partial x^2} , \quad -l \leq x \leq l ,
\end{equation}

\noindent with a reactive (Robin) boundary condition (i.e. Eq.~\eqref{eq:rxnBC1d}) at $x = \pm l$ and an initial condition of constant concentration i.e. $C(x,t=t_0)=C_0=$ const.
The problem can be redefined in non-dimensional terms as

\begin{equation}\label{eq:PureDiff1d-NonDim}
\begin{cases}
\dfrac{\partial C'}{\partial t'}=\dfrac{\partial^2 C'}{\partial x'^2} & -1 \leq x' \leq 1 \\
C'=1 & t=0\\
\Da\dfrac{\partial C'}{\partial x'}= \mp C' & x'=\pm 1
\end{cases}
\end{equation}

\noindent where $x'=x/l$, $t' = t D / l^2$, $C'=C/C_0$ and the Damk\"ohler number $\mathrm{Da}= k l / D$ represents the ratio between the characteristic diffusion time and the characteristic reaction time.
Note that in this section primes denote non-dimensional  variables, not to be confused with derivatives in Section \ref{sec:convergence_proof}.
The problem defined by~\eqref{eq:PureDiff1d-NonDim} has an analytical solution \citep[][section 3.11]{CJbook}, given by (see Fig.~\ref{fig:1D-CTimeSpace})

\begin{equation}\label{eq:CJsolution}
C' =\sum_{n=1}^\infty \dfrac{2 \rm{Da} \; cos\left(\alpha_n x'\right) sec(\alpha_\textit{n})}{\rm{Da}(\rm{Da}+1)+\alpha_\textit{n}^2} e^{-\alpha_\textit{n}^2t'}
\end{equation} 

\noindent where $\alpha_n$ are the positive roots of $\alpha \rm{tan}(\alpha) = \rm{Da}$.

%

In non-dimensional terms, the expression for the probability of reaction $p_1$ becomes $p_1=\rm{Da}$ $\sqrt{\pi \Delta t'}$, where $\Delta t' = D \Delta t / l^2$.
Random walk simulations with $p=p_1$ and $p=p_2$ were performed, distributing $N_{p,0}$ = 5$\times 10^6$ particles in the domain at the beginning of the simulation, and implementing the 1D random walk scheme.
For each time step the concentration in the domain was evaluated from the particles position distribution by discretizing the domain into a number of bins $N_{\rm{bins}}=50$, and calculating the normalized concentration $\hat{C_{i}}$ in each bin as

\begin{equation*}
\hat{C_{i}}=\dfrac{C_i}{C_0}=\dfrac{m_p(t) N_{p,i} N_{\rm{bins}}}{N_{p,0}m_{p,0}} \; .
\end{equation*}

\noindent where $m_p(t)$ is the mass of the particle and $m_{p,0}$ is arbitrarily set to 1.
The decrease in particle numbers over time lead to a decrease in the statistical significance of the result.
This problem was tackled by a mass-conserving splitting of the particles (see~ \ref{App:A} for details).

The results of normalized concentration values are illustrated in Fig.~\ref{fig:1D-CTimeSpace} for a specific $\Delta t' = 5 \times 10^{-3}$ and $\Da=1$, yielding $p_1=0.1253$ and $p_2=0.1179$ (a $\sim$~6\% difference).
As it can be seen qualitatively in Fig.~\ref{fig:1D-CTimeSpace}, the first order approximation for $p$ results in a noticeable underestimation of $\hat{C}_i$ everywhere in the domain, while the second order approximation leads to results much closer to the analytical solution.
The quantitative comparison between the analytical predictions and the random walk solution is conducted via two error metrics; $E_{\rm{norm}}$, the root mean square (RMS) of the normalized error and $E_{\rm{abs}}$, the RMS of the absolute error (see Appendix~B for definitions).

\begin{figure} 
	\begin{center}
		\includegraphics[width=0.9\textwidth]{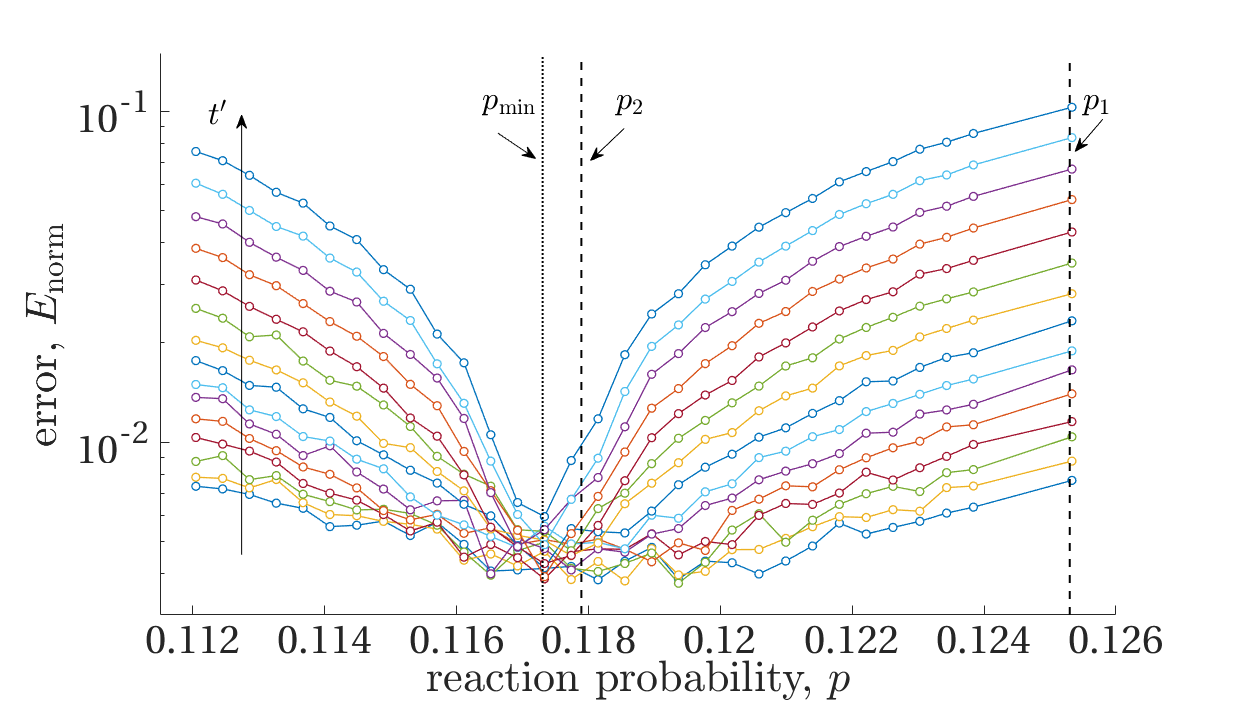}
		\caption{Values of normalized error of the 1D simulation as a function of the reaction probability. Here, $\Da = 1 $ and $\Delta t'=5\times10^{-3}$. 
			The dashed vertical lines indicate the first and second order approximations for $p$, respectively $p_1=0.1253$ and $p_2=0.1179$, while the dotted line points at the approximate global minimum of the normalized error,  $p_{\rm{min}}=0.1173$. The differently colored lines 
  	        correspond to times increasing in a geometric sequence, from $t'=0.1$ to $t'_{end}=3.05$.}
		\label{fig:PminSweep}
	\end{center}
\end{figure}

In Fig.~\ref{fig:1D-ErrorsSameDA}, we show the evolution with time of the errors $E_{\rm{norm}}$ and $E_{\rm{abs}}$ for the same Damk\"ohler number $\Da=1$, with two different time-steps $\Delta t'=\{5\times10^{-3},5\times10^{-4}\}$.
For each of the two cases three simulations were performed using reaction probabilities $p_1$, $p_2$, and $p_{\rm{min}}$.
\begin{figure} 
	\begin{center}
		\includegraphics[width=1\textwidth]{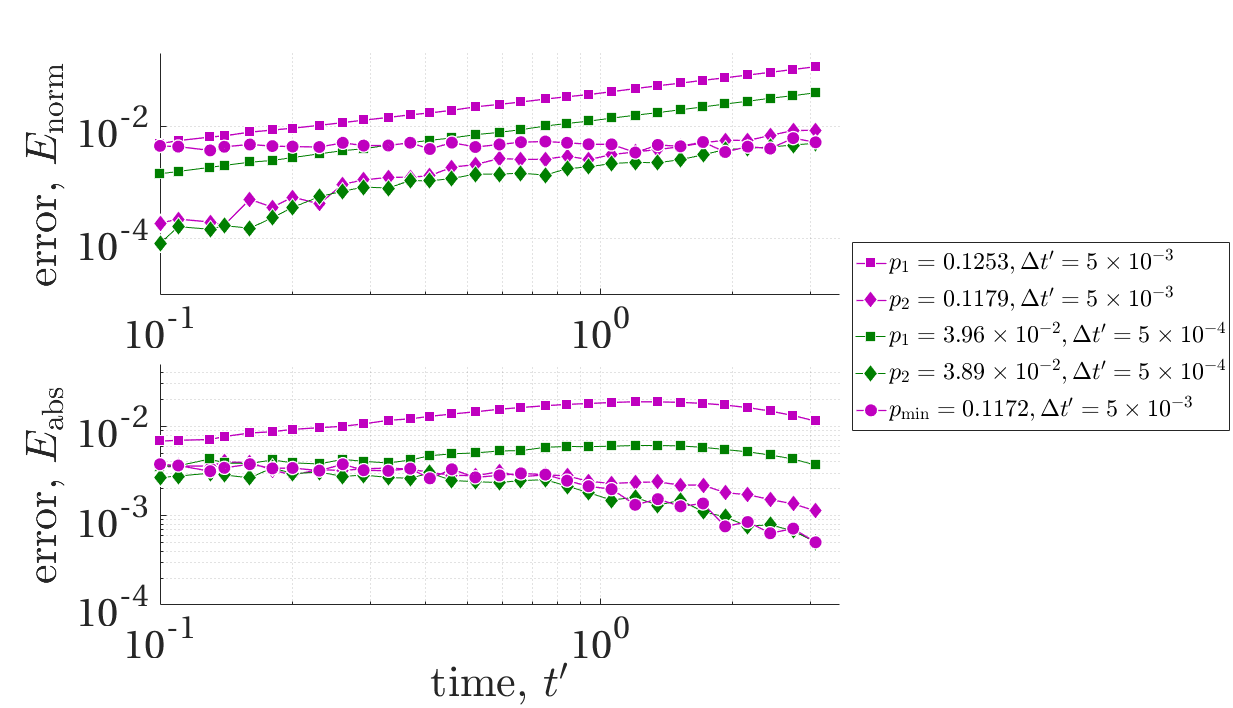}
		\caption{Normalized error $E_{\rm{norm}}$ (top) and absolute error $E_{\rm{abs}}$ (bottom) over time: values for simulations with $\Delta t'=5\times10^{-3}$ and $\Delta t'=5\times10^{-4}$ are shown (purple and green datasets, respectively; color in the online version of this paper).
			Different approximations for the boundary reaction probability $p$ are employed: $p=p_1$ (squares), $p=p_2$ (diamonds), $p=p_{\rm{min}}$ (circles).
			In all these simulations, Da=1 and $N_{p,0}=5\times10^6$.}
		\label{fig:1D-ErrorsSameDA}
	\end{center}
\end{figure}
The probability $p_{\mathrm{min}}$ minimizing this error was found numerically via a parametric sweeping operation: for the same case setup, a range of different probabilities was tested in the algorithm, and for each one, $E_{\rm{norm}}$ was calculated.
The results can be seen in Fig.~\ref{fig:PminSweep} where $E_{\rm{norm}}$ is shown as a function of $p$ for different times.

From this figure two points stand out: first, the steep descent of the error curves close to the global minimum shows how even minor differences in the probability used can lead to significant discrepancies from the expected concentration values, and second, it shows how simply implementing the second order correction (with respect to $p_1$) leads to an improvement in the error of almost one order of magnitude (for this case).
For the sake of clarity, we remind the reader that the difference between $p_{\rm{min}}$ and $p_2$ accounts for the higher-order terms we dropped from Eq.~\ref{eq:p2}, summed in $\epsilon$, meaning that $p_{\rm{min}}$ cannot be calculated analytically \textit{a priori}.
From these results in Fig.~\ref{fig:1D-ErrorsSameDA}, the argument just exposed is made even more clearly, as it can be seen that a much greater increase in the accuracy of the simulation can be gained by using the more precise second-order approximation for the estimation of $p$, with respect to the more immediate (but much costlier) solution of simply decreasing sharply the integration step $\Delta t'$.
The appreciable error in the lowest part of Fig.~\ref{fig:PminSweep} are attributable to the sampling error when binning a set of points.
This error is expected to be equal to $\sqrt{N_{\rm{bins}}/N_{p,0}}$ $\approx \sqrt{10^{-5}} \approx 3 \cdot 10^{-3}$, closely fitting the observed error.

\begin{figure} 
	\begin{center}
		\includegraphics[width=1.2\textwidth]{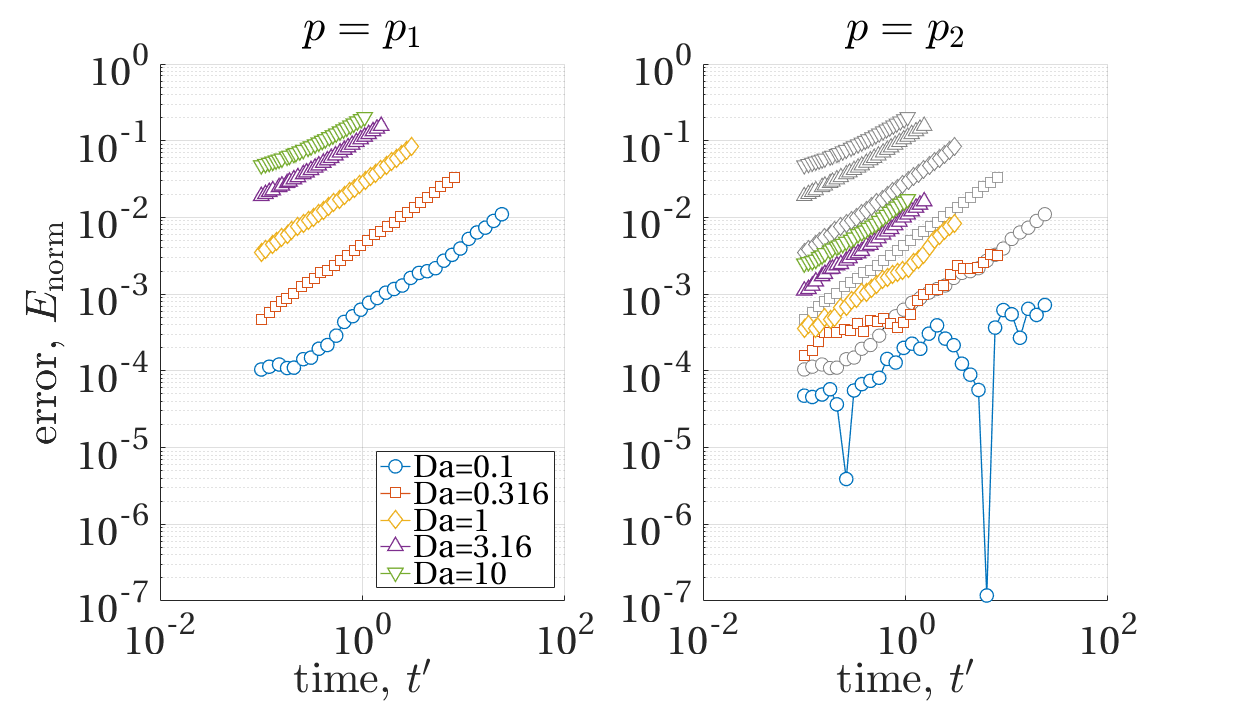}
		\caption{Normalized error $E_{\rm{norm}}$ over time, for $\Delta t' = 5\times 10^{-3}$, for five different Damk\"ohler numbers, equal to $\Da=0.1, 0.316, 1.0, 3.16,10$. On the left results for $p=p_1$ are shown (in color), while the figure on the right shows results for $p=p_2$ (in color) together with the same results for $p=p_1$ (in grey) for ease of comparison. } 
		\label{fig:ErrorsDiffDa}
	\end{center}
\end{figure}

Finally, simulations exploring the system behavior over a range of different Damk\"ohler numbers (see Tab.~\ref{tab:1D-DiffDa}) were performed.
\begin{table} 
	\centering
	\begin{tabular}{|c|c|c|}
		\hline
		Da & $p_1$ & $p_2$\\
		[1ex]
		\hline\hline
		0.10	&0.007927	&0.007895\\
		0.316	&0.02506	&0.02475\\
		1	    &0.07926	&0.07624\\
		3.16	&0.2506		&0.2227\\
		10		&0.7926		&0.5676\\
		\hline
	\end{tabular}
	\caption{Parameters for the simulation campaign shown in Fig.~\ref{fig:ErrorsDiffDa}. In all cases, $N_{p,0}=5\times10^6$ and $\Delta t'=2.5\times10^{-3}$.}\label{tab:1D-DiffDa}
\end{table}
Fig.~\ref{fig:ErrorsDiffDa} shows the evolution of relative error over time for the a few Da values. In this figure we compare the use of $p=p_1$ and $p=p_2$ in the random walk algorithm for a fixed $\Delta t' =2.5 \times 10^{-3}$ for all cases. The parameters of the cases (Da, $p_1$ and $p_2$) are given in Table \ref{tab:1D-DiffDa}.
Each case has a different end time $t'_{end}$ defined as the time when the average concentration becomes 10\% of the initial one, i.e., $\int_{-l}^{+l}(C / C_0)dx~=~0.1$.
As can be seen in this figure, the normalized errors for the $p_2$ case are smaller than the normalized errors for $p_1$. 
The error is reduced by about a factor of 2 in the early times in the smallest Da used (Da=0.1) and by more than an order of magnitude in the highest Da used (Da=10). 
We also note the gradual increase with time in all cases, which can be attributed to the use of a higher reaction probability $p$ with respect to the correct one, leading to a larger removal of particles each timestep and an increase in the mismatch between simulation and analytical solution.
Note that $p_1$ scales linearly with Da, so higher Da values correspond to higher $p_1$ and a more significant difference between $p_1$ and $p_2$.

For completeness and reference, the computational cost of the simulation showing the largest error in Fig.~\ref{fig:1D-ErrorsSameDA} (corresponding to $\Da=1$ and $\Delta t'=5\times10^{-3}$) was of about one minute, while the corresponding simulation with a smaller time-step discretization ($\Delta t'=5\times10^{-4}$) had a linearly increased cost of ~10 minutes. 
All of the simulation runs were performed using \verb+Matlab+ as a single-thread operation on a 2.6 GHz i7-6700HQ CPU.


\subsection*{Poiseuille flow between reactive parallel plates}
\begin{figure} 
	\begin{center}
		\includegraphics[width=0.7\textwidth]{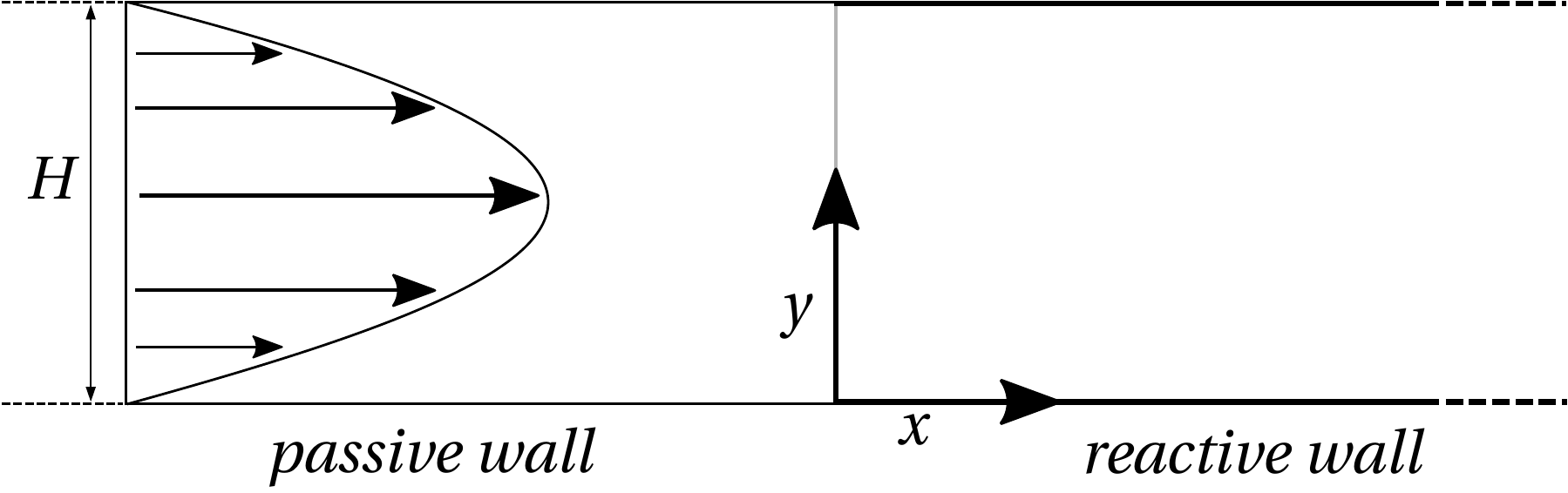}
		\caption{Setup of the 2D problem with Poiseuille flow between two plates.
			The plates are passive ($k=0$) at $x<0$ and reactive ($k>0$) at $x>0$.
			The reactive walls are denoted by thicker lines.}
		\label{fig:Sketch2D}
	\end{center}
\end{figure}

In this part of this work, we consider a problem of steady-state advection-diffusion-reaction in a 2D domain.
The geometry considered is an infinite strip, $\Omega=\{ -\infty <x<0, 0\leq y \leq H\}$ with width $H$, whose axis is parallel to the $x$ direction (see Fig.~\ref{fig:Sketch2D}).

Flow in the strip is assumed to be laminar and incompressible with a velocity profile $u_x(y)$ given by Poiseuille's law,

\begin{equation}\label{eq:Poiseuille}
u_x=\dfrac{H^2}{2\mu} \nabla p\left(\dfrac{y}{H}\left(1-\dfrac{y}{H}\right)\right) = 6\bar{u} \left( \dfrac{y}{H}\right) \left( 1-\dfrac{y}{H}\right) \quad 0\leq y\leq H
 \; ,
\end{equation} 

\noindent where $\mu$ is the fluid viscosity, $\nabla p$ the applied pressure gradient, and $\bar{u}$ is the average velocity; the fluid velocity has no transverse component, i.e. $u_y=0$.
The governing equation for the concentration $C$ is given by the ADE \eqref{eq:ADE}.
The walls of the domain ($y=\{0, H\}$) are assumed to be reactive in $x>0$ and passive in $x\leq0$.
This boundary condition is defined as $\mp D \frac{\partial C}{\partial y}=kC$, at $x>0, y=\{0, H\}$, where $k=\rm{const}>0$.
The concentration far upstream (at $x\rightarrow -\infty$) is assumed to be constant, $C=C_0$.

We employed the random walk algorithm to solve he problem and focused our attention on the steady state solution for this advection-diffusion problem.
To represent the boundary condition of constant concentration at $x\rightarrow -\infty$, a fixed number of new particles $N_{inj}$ was introduced every time step at the position $x_{inj}<0$.
The location $x_{inj}<0$ was chosen such that it is far enough upstream to represent the boundary condition, but not too far to avoid excessive computational burden; at the start of the RW simulation, there are no particles in the domain.
The code was run until convergence to steady state.
In non-dimensional terms, the steady state concentration $C'=\frac{C}{C_0}$ is a function of $x'=\frac{x}{H}$, $y'=\frac{y}{H}$, P\'eclet and Damk\"ohler numbers, i.e.: $C'=C'(x',y',\rm{Pe},\rm{Da})$, where Pe=$\bar{u} H / D$ and Da=$k H / D$.
We thus explored the problem for a range of Pe and Da.
Two simulation campaigns  were performed, each composed of nine different cases exploring the combination of Pe=1, 10, 100 and Da=1, 10, 100; as it was done in the 1D case, the two campaigns differed in the use of the approximation for reaction probability: $p=p_1$ or $p=p_2$, respectively (details in Tab.~\ref{tab:2D-Setup}).

\begin{table} 
\centering
\begin{tabular}{|c|c||c|c|c|}
\hline
Pe& Da & $N_{inj}$& $p_1$ & $p_2$\\
[1ex]
\hline\hline
1	&1	&5   &0.01 & 0.00995  \\
1	&10	&5   &0.1  & 0.0952 \\
1	&100&5   &1    & 0.667 \\
10	&1	&17  &0.01 & 0.00995 \\
10	&10	&17  &0.1  & 0.0952 \\
10	&100&17  &1    & 0.667 \\
100 &1	&17  &0.01 & 0.00995 \\
100 &10	&17  &0.1  & 0.0952 \\
100 &100&17  &1    & 0.667 \\
\hline
\end{tabular}

\caption{Parameters for the 2D simulation campaign, showing $N_{inj}$, $\Delta t$, and $p$ for all cases.
In every case, $\Delta t'=3.18\times10^{-5}$, where $\Delta t'=\Delta t \dfrac{D}{H^2}.$}
\label{tab:2D-Setup}
\end{table}

\begin{figure} 
	\begin{center}
		\includegraphics[width=1\textwidth]{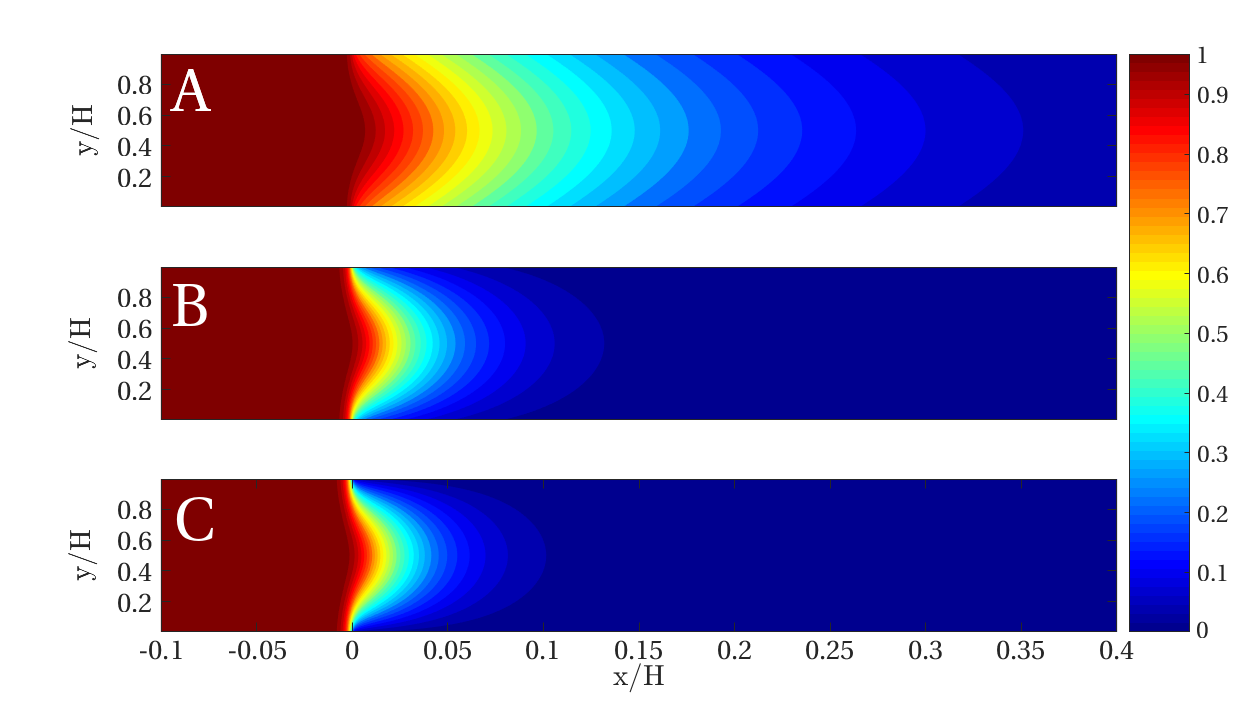}
		\caption{Solution of non-dimensional particle concentration $C/C_0$ in the 2D from Comsol, for Pe=10 and Da=1, 10, 100 at steady-state (respectively, cases A, B, and C).
			The reactive walls are present at $x>0$; at $x<0$ it is possible to notice particle backdiffusion, especially at higher Damk\"ohler numbers (bottom).
			It has to be noted that in this figure, the domain aspect ratio has been skewed for the sake of visualization clarity.}
		\label{fig:Contours}
	\end{center}
\end{figure}

An analytical solution for the problem at hand is not available.
Thus, in order to test the performance of the random walk algorithms, we performed additional Eulerian simulations on the same cases explored in the Lagrangian code, using the finite element commercial CFD suite \verb+Comsol 5.2a+.
The velocity field and the concentration in the domain for each of the nine cases were obtained by first solving the laminar flow problem, and then the steady-state form of the advection-diffusion equation (i.e. Eq.~\eqref{eq:ADE} with the time derivative set to zero), with the partially reacting boundary conditions on the walls.
For illustration, we show in Fig.~\ref{fig:Contours} the results of three of the nine cases (Pe=10, Da=1, 10, 100): note (especially in the bottom figure) that the normalized concentration is not uniform, and is smaller than unity close to the boundaries.
This is due to back-diffusion from the reactive area of the domain, $x>0$.
In order to compare the RW and \verb+Comsol+ results, we calculated the total mass in the semi-infinite $x>0$ domain, 

\begin{equation}
M'=\int_0^1 \int_0^{\infty} C' dx' dy' \;
\end{equation}

\begin{figure} 
	\begin{center}
		\includegraphics[width=.9\textwidth]{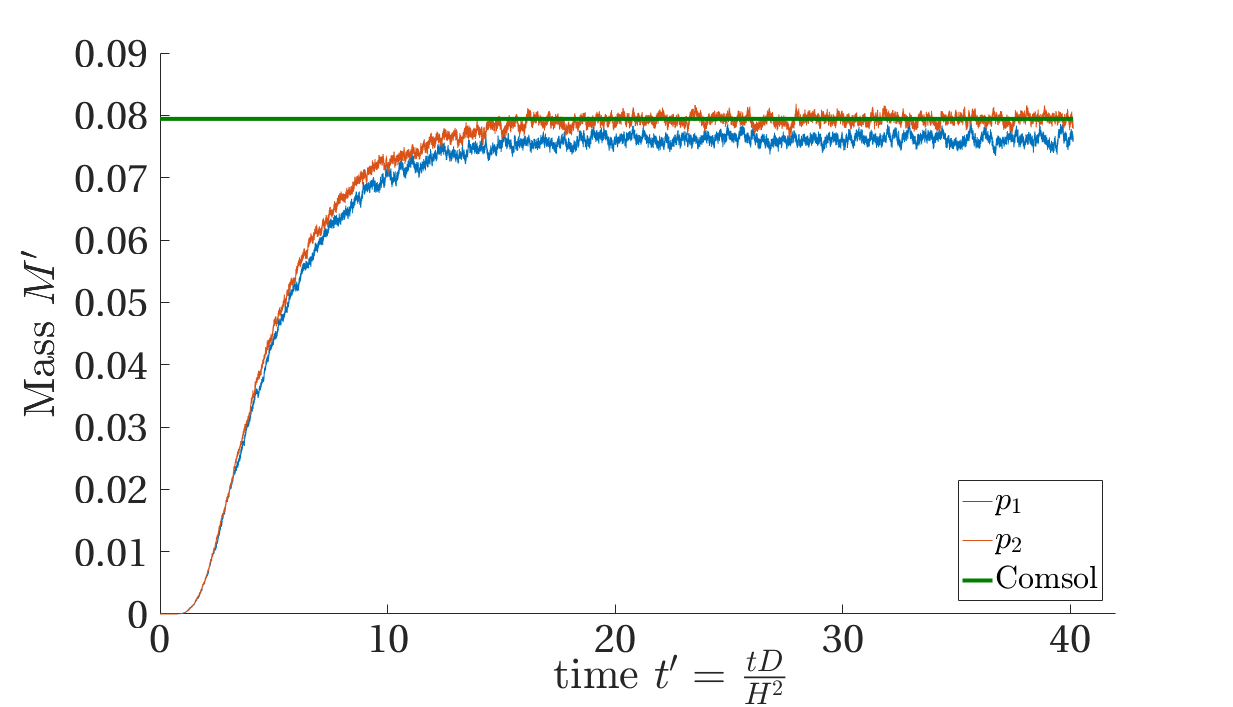}
		\caption{Time evolution of total mass $M'$ at $x>0$ for the random walk code, comparing first order (blue) with second order approximation (red) for reaction probability $p$. Steady state result of the Comsol simulation is shown for comparison, represented as a straight line.
			Data shown for Pe=100, Da=1 case.}
		\label{fig:2D-MassOverTime}
	\end{center}
\end{figure}

Figure~\ref{fig:2D-MassOverTime} shows the evolution in time of $M' $ for a specific case (Pe=100, Da=1).
In this figure, it is qualitatively clear that there is an improvement of the random walk algorithm accuracy when using the second order approximation; in the other cases it is hard to visually discern the lines corresponding to $p_1$ and $p_2$, so for the sake of brevity we do not include similar figures for these cases here.
In order to provide a more quantitative measure for the mismatch, we calculate the relative error by comparing RW results and the \verb+Comsol+ simulations (see \ref{App:B}): this data is shown in Figure~\ref{fig:2D_error_Pe_DA}.
To provide an accurate estimation of the uncertainty of the \verb+Comsol+ results, we performed a grid-convergence campaign for each of the nine explored cases, then calculated the \textit{Grid Convergence Index} using the well-known Richardson extrapolation procedure, later improved by Roache~\cite{roache1998fundamentals,roache1998verification} (see \ref{App:C}).
The first point to note is that, for all cases, the error never exceeds a few percentages, showing the accuracy of the random walk code (or at very least, the closeness of the results to the estimate from the Eulerian data).
Clearly, a marked improvement in the code accuracy was obtained by simply employing the second order approximation for the reaction probability.
The figure also shows the standard deviation of the presented data, showing the uncertainty due to the calculation of steady-state average particle concentration (due to small oscillations in particles number, see Fig. \ref{fig:2D-MassOverTime}) and the remaining uncertainty in the \verb+Comsol+ results (see~\ref{App:C}).
As it is apparent, the error bars show how the predictions from the two different probability estimation strategies can be reliably set apart, and the difference be ascribed to a fundamental improvement in the code accuracy.
While there is no discernible trend linking the calculated error with the problem setup (e.g.: P\'eclet and Damk\"ohler number), in all cases the more accurate campaign using $p=p_2$ shows errors ranging from ~$10^{-4}$ to ~$10^{-3}$. 

Lastly, the runtime of the slowest PT simulation in this test case (Da=1) was equal to $\sim$3 hours per simulation for Pe=10 and Pe=100, and $\sim$30 hours for Pe=1.
In comparison, the runtime of the Comsol steady state model ranged between a few minutes to $\sim$1 hour per simulation (for the same setups, with higher cost for higher P\'eclet numbers).
The Comsol simulations were run in parallel on an esa-core Intel Core i7-3960x workstation; while PT simulations were run on an Intel Xeon  E5-2630 on a single thread.
The difference is attributed first on the parallelization difference, and also to the slower convergence of the transient RW code to an apparent steady state.


\section{Conclusions}
In this work we treated the implementation of the partially reacting boundary condition in random walk numerical algorithms.
We start from a simple relationship between the reaction probability $p$ and the algorithm time step $\Delta t$, employed by previous works, where $p \propto \sqrt{\Delta t}$.
This relationship is correct at first order.

First, we give a theoretical analysis resulting in an estimation for $p$ correct to the second order, and an estimation of its third-order error.
Then, in order to show the increase in prediction accuracy which can be gained from this correction, we set up two different test cases and compare the effectiveness of the classic methodology with the one proposed in this work.
In the first case we study a simple 1D pure diffusion problem, for which an analytical solution is available.
We use this analytical solution to calculate the error in the random walk predictions under a wide range of reaction rates.
We observe that while the classical first order approximation described the system in a qualitatively satisfactory manner, the use of the proposed second order approximation for the reaction probability reduced the RW simulation error by about an order of magnitude for the cases considered.
Similar results are obtained when studying a more realistic 2D advection-diffusion problem: in this case a wide parametric sweep over P\'eclet and Damk\"ohler numbers was performed.
The results were compared with a finite element steady state simulation.
Again we observe that the use of the second order approximation for $p$ improves the solution accuracy over all the explored parameter space, proving the reliability and effectiveness of the proposed methodology.

The main result of this work is thus to provide a simple way of improving the predictive capabilites of Lagrangian random walk algorithms in the case of reacting boundary conditions.
Such a reactive boundary is a very common physical setup and of great interest in the modelling of a diverse range of applications both in reaction engineering and environmental science.
This kind of improvement should afford practitioners a greater reach when dealing with the trade-off between simulation accuracy and computational cost.

\section*{Acknowledgements}
A.P. and G.B. would like to acknowledge partial funding by the Startup grant of the V.P. of research in Tel Aviv University and partial funding by the Israel Water Authority.


\begin{figure} 
\begin{center}
\includegraphics[width=.85\textwidth]{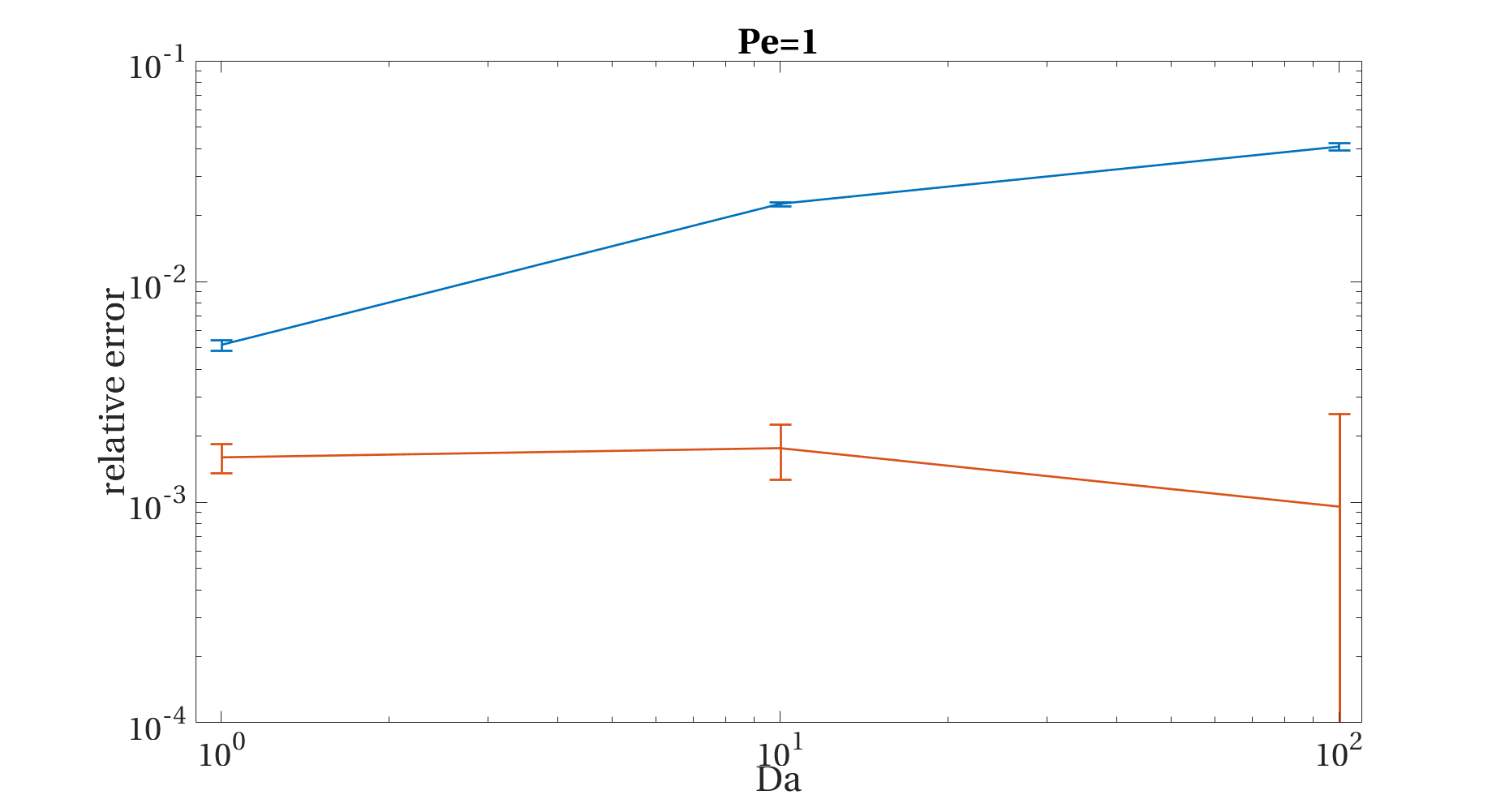}
\includegraphics[width=.85\textwidth]{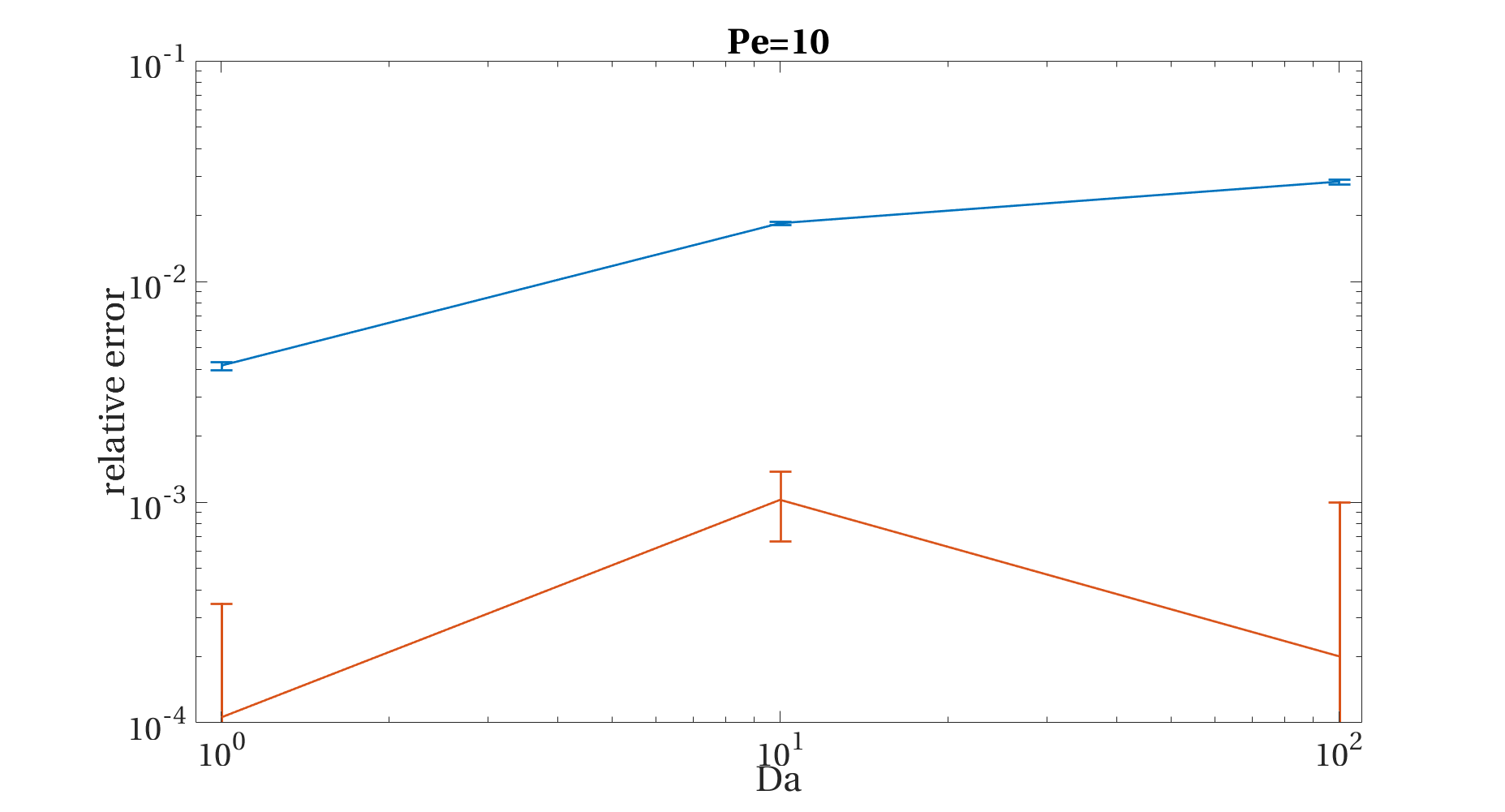}
\includegraphics[width=.85\textwidth]{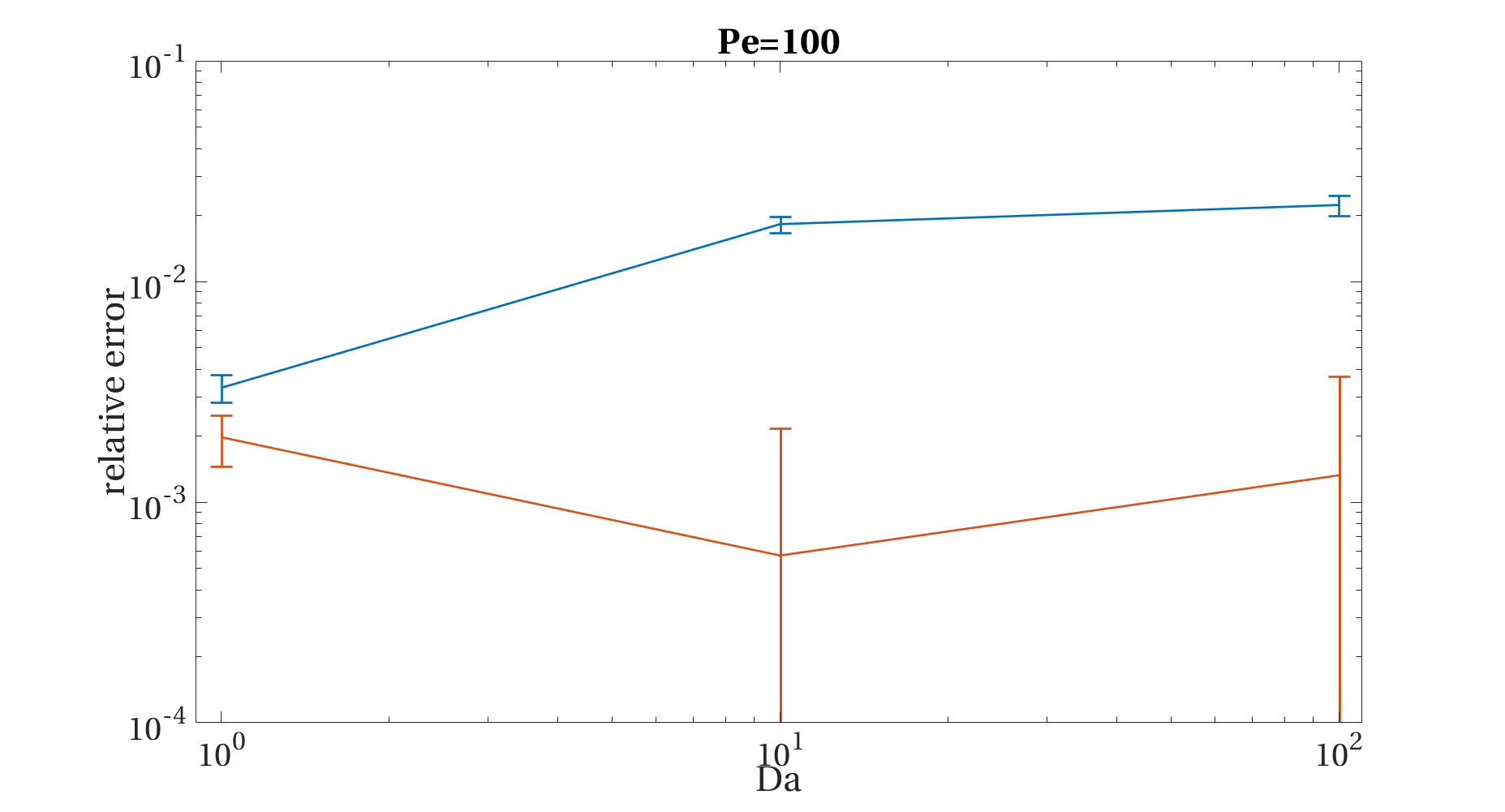}

\caption{Relative error between random walk simulations results and grid-converged Eulerian data, shown for the range of P\'eclet and Damk\"ohler numbers. Results from the $p=p_1$ (blue) $p=p_2$ (red) campaigns are compared. The error bars show the uncertainty in the calculation owing to particle number oscillations in the random walk code and the estimated errors in the simulations.}
\label{fig:2D_error_Pe_DA}
\end{center}
\end{figure}

\clearpage
\section*{References}
\bibliographystyle{model3-num-names}
\bibliography{biblio}

\appendix
\section{Treatment of particle number evolution in time}\label{App:A}
\noindent It has to be noted that in the 1D case the number of particles will monotonically decrease in time, impairing the validity of the statistics at late times, as the number of remaining particles in the system approaches zero.
In order to balance out this problem, we implemented a numerical fix in our algorithm: when the number of particles became smaller than half of the initial number, they were split into two new particles, each of which now carrying half the mass of the parent particle.
Then, all the relevant statistics (e.g.: density functions) were calculated based on the particles updated mass.
Note that mass $m_p(t)$ is uniform for all particles.

\section{Normalized and absolute errors} \label{App:B}

\noindent When comparing the random walk results with the analytical solution for the 1D case, the root mean square of the normalized and absolute errors are employed as the relevant metrics.
The normalized error is defined as

\begin{equation}\label{eq:normRMSerror}
E_{\rm{norm}}=\sqrt{\dfrac{\sum_{n=1}^{N_{\rm{bins}}}\left( \dfrac{C_{i}-\hat{C_i}}{C_{i}}\right)^2}{N_{\rm{bins}}}} \; ,
\end{equation}

\noindent where $C_{i}$ is the analytical solution of Eq.~\eqref{eq:CJsolution} evaluated at each bin mid-point.
The absolute error is defined as

\begin{equation}\label{eq:absRMSerror}
E_{\rm{abs}}=\sqrt{\dfrac{\sum_{n=1}^{N_{\rm{bins}}}\left( C_{i}-\hat{C_i}\right)^2}{N_{\rm{bins}}}} \; ,
\end{equation}

Similarly, when treating the results of the 2D RW simulations, we calculated the relative error between RW and \verb+Comsol+ as:

\begin{equation}\label{eq:normRMSerror2D}
E_{\rm{norm},2D}=
\bigg\lvert
\dfrac{M'_{RW} - M'_{\textsc{Comsol}}}{M'_{\textsc{Comsol}}} 
\bigg\rvert 
\;.
\end{equation}

\section{Grid convergence analysis} \label{App:C}

Any numerical simulation is prone to suffer from some measure of error: these could come from an incorrect modelling or (in the case of Eulerian simulations) from an insufficient discretization of the computational mesh.
In order to minimize the latter, the usual procedure is to perform a grid independence study, by successively refining the mesh until there is a reasonable certainty that the discretization errors are smaller than the desired accuracy.

In our case we aim to use Eulerian simulation results to evaluate the accuracy of our PT runs, and we do not possess either analytical or empirical validation data sets accurate enough to distincly discern the effects of using $p_1$ or $p_2$.
Specifically, the remaining discretization error even after a grid convergence study could be of the same magnitude of the difference between $p_1$ and $p_2$ PT results, making any comparison moot.
In this work, we calculated the uncertainty of our \verb+Comsol+ simulation results by using a methodology based on the Richardson extrapolation~\cite{richardson1927}, expanded on by Roache~\cite{roache1998verification,roache1998fundamentals}.

In short, the basic assumption is that the discrete solutions $f$ obtained with a numerical method can have a series representation

\begin{equation}\label{eq:richardson}
f=f_{exact}+g_1h+g_2h^2+g_3h^3+\dots
\end{equation}

where $f_{exact}$ is the exact solution, $h$ is the grid spacing employed, and the functions $g_i$ are defined in the domain and do not depend on the discretization~\cite{roache1998fundamentals}.
Various methods for the calculation of the \textit{Grid Convergence Index} (GCI) and the estimation of $f_{exact}$ have been used, both for finite-volume and finite-element analyses~\cite{kwasniewski2013}.
In this work we followed the procedure reported in Celik~\cite{celik2008}, used also in very recent works~\citep{mansour2018}.
For brevity, we refer the reader to the former very clear and succinct reference for the full rundown of the method, complete with step-by-step instructions: we will just give a very brief exposition here.
In our case, we performed a simulation campaign akin to an usual grid convergence study for each of the nine cases explored, with successively smaller grid spacings $h_3$, $h_2$, and $h_1$.
Then, we calculated $\phi_3$, $\phi_2$, and $\phi_1$, namely the average concentration values from each subsequent grid refinement.
These were used to estimate the \textit{order of convergence}, $p$ (for the procedure and the formula we refer again to~\cite{celik2008}).
Together with the grid refinement factor $r$ (i.e. the ratio between successive grid spacings), the GCI can be calculated as:

\begin{equation}\label{eq:GCI}
GCI=F_s\dfrac{|\varepsilon|}{r^p-1} \;,
\end{equation}

\noindent where $\varepsilon$ is the relative error between the two most refined grids (i.e.: $\frac{\phi_1-\phi_2}{\phi_1}$) and $F_s$ a safety factor, set equal to 1.25 when working with three or more meshes (as in our case).
The resulting GCI, expressed as a percentage, can be considered as a relative error bound showing how the solution calculated for the finest mesh is far from the asymptotic value~\cite{roache1998verification}.
This value (calculated for each of the nine cases), together with the numerical uncertainty of the RW simulations, constitutes the total uncertainty of the results, and is represented in the error bars in Fig.~\ref{fig:2D_error_Pe_DA}.


\end{document}